\documentclass[doublecol]{epl2}

\title{
Nucleation process in the Burridge-Knopoff model of earthquakes
}

\author{
Y. Ueda, S. Morimoto, S. Kakui, T. Yamamoto \and H. Kawamura
}

\institute{                    
Department of Earth and Space Science, Faculty of Science,
Osaka University, \\
Toyonaka 560-0043, Japan
}


\abstract{
Nucleation process of the one-dimensional Burridge-Knopoff model of earthquakes  obeying the rate- and state-dependent friction law is studied both analytically and numerically. The properties of the nucleation dynamics, the nucleation lengths and the duration times are examined together with their continuum limits.
}

\begin{document}

\maketitle

There is a wide-spread expectation that a large earthquake might be preceded by a precursory nucleation process which occurs prior to the high-speed rupture of a mainshock. Nucleation process is localized to a compact ``seed'' area with its rupture velocity orders of magnitude smaller than the seismic wave velocity \cite{Dieterich09,Scholzbook,Dieterich92,Ohnaka}. The fault spends a very long time in this nucleation process, and then at some point, exhibits a rapid acceleration accompanied by a rapid expansion of the rupture zone, finally getting into the high-speed rupture of a mainshock. Such a precursory phenomenon preceding a mainshock is of paramount importance in its own right as well as in its possible connection to an earthquake forecast.

 It has been suggested that the earthquake nucleation process might proceed via several distinct steps or ``phases''. Ref.\cite{Ohnaka} proposed that it started with an initial quasi-static process until the nucleus diameter $L$ exceeded a nucleation length $L_{sc}$. Then, the fault gets into the acceleration phase where the system gets out of equilibrium and rapidly increases its slip velocity. When the nucleus diameter exceeds another nucleation length $L_{c} (> L_{sc})$, the fault eventually exhibits a high-speed rupture of a mainshock. In this picture, two nucleation lengths, $L_{sc}$ and $L_{c}$, divide the nucleation process into ``the initial phase''($L<L_{sc}$), ``the acceleration phase'' ($L_{sc}<L<L_c$) and ``the high-speed rupture phase'' ($L>L_c$). Although such features of the nucleation process have been more or less confirmed by laboratory rock experiments \cite{Latour,McLasky}, its nature, or even its very existence, remains less clear for real earthquakes \cite{Kato,Bouchon}.

 Under such circumstances, a theoretical or a numerical study based on an appropriate model of an earthquake fault would be important and helpful. In such modelings, the friction force is a crucially important part. The friction force now standard in seismology is the so-called rate and state dependent friction (RSF) law \cite{Dieterich79,Ruina,Marone}. The RSF law has been used in many of numerical simulations on earthquakes, mostly in the continuum model \cite{TseRice,Rice,Kawamura-review}, including earthquake nucleation process \cite{AmpueroRubin}.

 Meanwhile, a further simplified {\it discrete\/} model has also been used. Especially popular is the spring-block model or the Burridge-Knopoff (BK) model \cite{BK,CL,CLS-review,MoriKawamura05,MoriKawamura08,Kawamura-review,MoriKawamura08b}, in which an earthquake fault is modeled as an assembly of blocks mutually connected via elastic springs which are subject to the friction force and are slowly driven by an external force mimicking the plate drive. While a simple velocity-weakening friction has often been assumed in many numerical simulations, a more realistic RSF law was also employed in some of the recent numerical simulations \cite{CaoAki,OhmuraKawamura,Kawamura-review}. The model might also be useful in describing other stick-slip-type phenomena such as landslides \cite{Viesca}.

 The aim of the present letter is to clarify the nature of the nucleation process of the BK model. The BK model has widely been used especially in statistical physics, and clarifying the nature of its nucleation process would be important. Meanwhile, one may wonder what is the merit of studying the discrete BK model possessing an intrinsic  short-length cut-off in the form of block size, even though a similar nucleation problem was already studied within the continuum model \cite{Dieterich92,AmpueroRubin}. There might be two reasons for this. First, the issue of the discreteness is in fact closely related to the nucleation phenomenon. Rice argued that the characteristic length scale to be compared with the block size was the nucleation length, and the continuum system under the RSF law always exhibited a nucleation process prior to a mainshock \cite{Rice}. We wish to clarify how the nucleation process of the discrete BK model behaves in its continuum limit, by systematically varying the extent of the discreteness of the model. Note that the extent of the discreteness may be regarded as a measure of the underlying spatial inhomogeneity \cite{Rice}. The second reason is more technical, {\it i.e.\/}, the BK model is much simplified compared to the continuum model, and often makes statistically relevant simulations possible in which hundreds of thousands of events are generated.

The one-dimensional (1D) BK model consists of a 1D array of $N$ identical blocks of the mass $m$, which are mutually connected with the two neighboring blocks via the elastic springs of the srping stffness $k_c$, and are also connected to the moving plate via the springs of the spring stiffness $k_p$, and are driven with a constant rate $\nu'$. All blocks are subject to the friction strength $\Phi$. The equation of motion for the $i$-th block can be written as \cite{Kawamura-review}
\begin{equation}
m \frac{{\rm d}^2U_i}{{\rm d}t^{\prime 2}} = k_p (\nu ' t'-U_i) + k_c (U_{i+1}-2U_i+U_{i-1})-\Phi_i,
\label{original-eq-motion}
\end{equation}
where $t'$ is the time, $U_i$ is the displacement of the $i$-th block. This equation can be made dimensionless as

\begin{eqnarray}
\frac{d^2u_i}{dt^2} = \nu t-u_i+l^2(u_{i+1}-2u_i+u_{i-1}) - \phi_i ,
\label{eq-motion}
\end{eqnarray}
where $l \equiv (k_c/k_p)^{1/2}$. The dimensionless displacement $u_i$ is normalized by the critical slip distance ${\mathcal L}$ associated with the RSF law, the time $t$ by $\omega^{-1}=\sqrt{m/k_p}$, the block velocity $v_i$ and the pulling speed of the plate $\nu$ by $\mathcal{L}\omega$, and the dimensionless friction force $\phi$ by $k_p{\mathcal L}$. The RSF force $\phi$ reads as
\begin{equation}
\phi_i=c+a\log(1+\frac{v_i}{v^*})+b\log \theta_i ,
\label{RSF}
\end{equation}
where $\theta_i$ is the dimensionless state variable describing the ``state'' of the interface, and $v^*$ is the dimensionless crossover velocity. The normalized frictional parameters $a$, $b$ and $c$ represent the velocity-strengthening, the velocity-weakening and the constant parts of friction. The original friction parameters $A$, $B$ and $C$ are related to the normalized ones by $A=(k_p{\mathcal L}/{\mathcal N})a$, $B=(k_p{\mathcal L}/{\mathcal N})b$ and $C=(k_p{\mathcal L}/{\mathcal N})c$, ${\mathcal N}$ being the normal load. For simplicity, we inhibit the  motion in the direction opposite to the plate drive. The state variable $\theta_i$ is assumed to obey the aging law \cite{Ruina},
\begin{equation}
\frac{{\rm d}\theta_i}{{\rm d}t} = 1-v_i\theta_i .
\label{aging}
\end{equation}

In the simplest version of the BK model as studied here, the nearest-neighbor interaction is assumed between blocks, while in real faults the crust perpendicular to the fault plane mediates the effective long-range interaction between blocks away on the fault plane. Indeed, such a long-range interaction was assumed in some of the previous studies of the BK model, especially its statistical properties such as the magnitude distribution \cite{MoriKawamura08b}. In the present study, however, we concentrate on the nucleation process of the nearest-neighbor model, with the aim of clarifying the nucleation process of the simplest version of the model.

 What type of setting the BK model actually assumes  in terms of an earthquake fault embedded in the 3D continuum crust might not be so trivial. The authors' view is as follows. Consider first the 2D BK model mimicking a planar fault embedded in the 3D continuum crust. Let the dimension of the block be $D\times D^\prime \times W$, where $D$ is the dimension along the plate drive, $D^\prime$ the dimension perpendicular to the plate drive within the fault plane, and $W$ the dimension perpendicular to the fault plane. Then, the block assembly represents a deformable ``fault layer'' of the width $W$ which is uniformly pulled by the more or less rigid plate contingent to it. Our estimate to be given below entails the width $W$ of order $\sim 2$ [km]. Thus, in the BK model, a uniform plate drive is applied not at infinity as boundary conditions as often assumed in the continuum model, but is applied rather close to the fault plane of order the distance $W\simeq 2$ [km]. Such a direct plate drive yields a term proportional to the displacement $-u_i$ in the equation of motion (2), which is absent in the standard elasto-dynamic equation. The 1D BK model is a simplification of the 2D model where one direction of the fault plane perpendicular to the plate drive has been integrated out, or supposed to be completely rigid.

 We note that the fault layer as modeled by the BK model might be related to the so-called ``low-velocity fault zones (LVFZ)'' observed in most mature faults, with $20\% \sim 60\%$ wave-velocity reduction relative to the host rock \cite{Huang}. Their widths were reported to be 100 [m] $\sim$ 2 [km], which are a bit smaller than, but does not much differ from the present estimate of the fault-zone width $W$.

 Let us try to estimate typical values of the model parameters with natural earthquake faults in mind. The dimensionfull rise time of an event, {\it i.e.\/}, the time elapsed from a given block involved in a mainshock rupture begins to move  until it stops, is found to be $\simeq \omega^{-1}$. This is true for a single-block system, while our simulations indicate it is also the case for a many-block system. Since the typical rise time of an earthquake is a few seconds, we get an estimate of $\omega ^{-1}\simeq 1$ [s].  The reported values of the critical slip distance ${\mathcal L}$ are largely scattered in the literature depending on the observation scale \cite{Scholz,Scholzbook,Toro}. Here, from our numerical observation that the typical block sliding velocity at the mainshock rupture is $10^2\sim 10^3$ in units of ${\mathcal L}\omega$ while it is around 1 [m/s] in real seismicity, we take ${\mathcal L}$ to be a few [cm], which is not far from the value at the seismic depth deduced in \cite{Scholz,Scholzbook,Toro}. Since the speed of the plate motion is typically a few [cm/year], the dimensionless loading rate is $\nu \simeq 10^{-7}-10^{-8}$.

The spring constant $k_p$ may be related to the rigidity $G$ as $k_p=G \frac{DD^\prime}{W}$. This can be derived by noting that the shear force $F_{shear}$ acting on a block with the displacement $U$ is given by $F_{shear}=k_pU=DD^\prime \times G\frac{U}{W}$ where $\frac{U}{W}$ is the shear strain. The relation $k_p=m\omega^2=\rho WDD^\prime \omega^2$ ($\rho$ is the mass density) and the $s$-wave velocity $v_s=\sqrt{\frac{G}{\rho}}$ yield $W=\sqrt{\frac{G}{\rho}}\frac{1}{\omega}=\frac{v_s}{\omega}$. Putting $v_s\simeq 2$ [km/s], which is taken somewhat smaller than the standard value of $v_s\simeq 3$ [km] due to the possible lower wave-velocity in the fault zone,  and $\omega^{-1}\simeq 1$ [s], we get an estimate of $W\simeq 2$ [km] as given above. The proportionality between the fault-zone width $W$ and the rise time $\omega^{-1}$ obtained here seems consistent with the observation on the LVFZ \cite{Huang}. With ${\mathcal N}=\sigma_nDD^\prime$ where $\sigma_n$ is the normal stress, we have $\frac{{\mathcal N}}{k_p{\mathcal L}}=\frac{\sigma_n v_s}{G \omega{\mathcal L}}$. Putting $\frac{\sigma_n}{G} \simeq 10^{-3}$, we get $\frac{{\mathcal N}}{k_p{\mathcal L}}\simeq 10^2-10^3$. As $C$ is known to take a value around $\frac{2}{3}$ \cite{Scholzbook}, $c$ would be of order $10^2$-$10^3$, $a$ and $b$ being one or two orders of magnitude smaller than $c$. The crossover velocity $V^*$ and its dimensionless counterpart $v^*$ is hard to estimate though it should be much smaller than unity, and we take it as a parameter in our simulations.

 The continuum limit of the BK model corresponds to making the dimensionless block size $d$, defined by $d=\frac{D}{v_s/\omega}$, to be infinitesimal $d\rightarrow 0$, simultaneously making the system infinitely rigid $l\rightarrow \infty$ with $d=1/l$ \cite{MoriKawamura08}. The dimensionless distance $x$ between the block $i$ and $i^{\prime}$ is $x=|i-i^{\prime}|d=\frac{|i-i^{\prime}|}{l}$. Notice that the continuum limit considered here concerns only with the fault direction (the fault plane in case of 2D), and the perpendicular direction ($W$-direction) is kept fixed. Thus, the possible internal motion in the fault layer along the perpendicular direction is suppressed in the model setting.

 As discussed in Ref.\cite{MoriKawamura08}, the 1D equation of motion in the continuum limit is given in the dimensionful form by
\begin{equation}
\frac{{\rm d}^2U}{{\rm d}t'^2} = \omega^2(\nu't'-U) + v_s^2 \frac{{\rm d}^2U}{{\rm d}x^2} - \Phi' ,
\label{continuumeq}
\end{equation}
where $U(x,t')$ is the displacement at the position $x$ and the time $t'$, $\Phi'$ is the friction force {\it per unit mass\/}, while $\omega$ and $v_s$ are the characteristic frequency and  the characteristic wave-velocity ($s$-wave velocity), respectively. As mentioned, the term $-\omega^2U$ representing the plate drive is absent in the standard elasto-dynamic equation. If one discretize the space into blocks of the size $D$ with $\frac{{\rm d}^2U}{{\rm d}x^2}\approx (U_{i+1}-2U_i+U_{i-1})/D^2$ and notes the relation $k_p=m\omega^2$ and $(v_s/D)^2=(d\omega)^2=(\frac{1}{l}\omega)^2$, one gets Eq.(1) with $k_c=l^2k_p$.

\begin{figure}
\onefigure[scale=0.8]{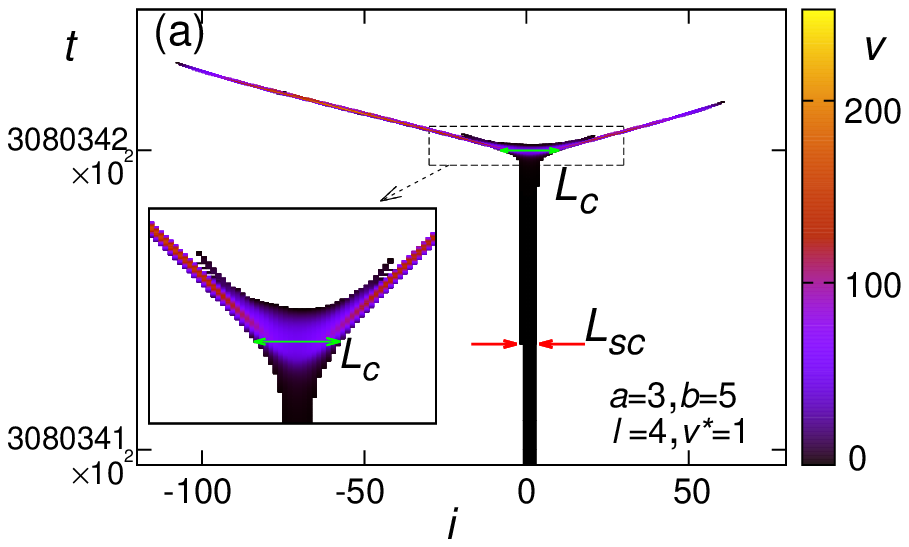}
\onefigure[scale=0.8]{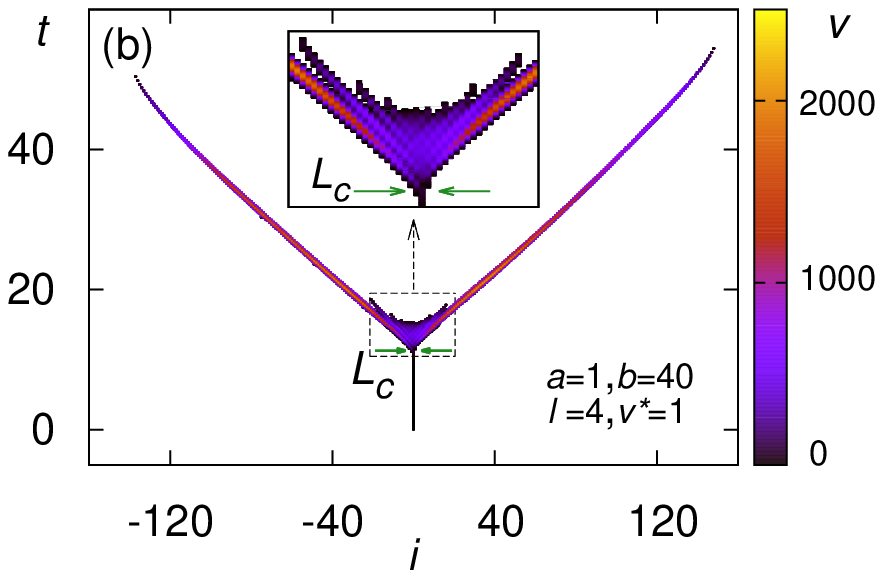}
\caption{
Color plots of typical earthquake nucleation processes depicted in the block-number (position) versus the time plane, (a) in the weak frictional instability regime, and (b) in the strong frictional instability regime. The color represents the block sliding velocity (white means exactly zero slip). The parameters are $a=3$, $b=5$ in (a), and $a=1$, $b=40$ in (b), with $c=1000$, $l=4$, $v^*=1$ and $\nu=10^{-8}$ in common. The origin of time ($t=0$) is taken to be the onset of the nucleation process where an epicenter block begins to move.
}
\end{figure}

 Our first question might be whether the 1D BK model under the RSF law ever exhibits a nucleation process prior to a mainshock, and if it does, under what conditions. We observe that the model exhibits qualitatively different behaviors depending on whether the frictional instability is either ``weak'' or ``strong''. A slow and long-lasting nucleation process, the quasi-static initial phase, is realized in the former case only. We illustrate in fig.1 typical examples of seismic events realized in the stationary state of the model, where the time evolution of the movement of each block is shown as a color plot for each case of (a) the weak frictional instability ($a=3$, $b=5$, $l=4$), and (b) the strong frictional instability ($a=1$, $b=40$, $l=4$). A slow nucleation process with a long duration time is observed in (a), but is absent in (b). As will be shown below, the model possesses a borderline value of $b$ determined solely by the stiffness parameter $l$, $b_c(l)=2l^2+1$, which discriminates the strong/weak instability behaviors.

 We also illustrate in fig.1 the two types of nucleation lengths, $L_{sc}$ and $L_{c}$ ($L_{sc} < L_c$). The former $L_{sc}$ is the length separating stable and unstable ruptures and exists only for the weak frictional instability, while the latter $L_{c}$ is the length signaling the onset of the high-speed rupture of a mainshock.

 One way to identify $L_{sc}$ is to artificially stop the external loading in the course of simulation. We have confirmed that, if the external loading is stopped at any point beyond $L=L_{sc}$, the subsequent seismic rupture is no longer stoppable and evolves until its very end, whereas, if the external loading is stopped at a point before $L=L_{sc}$, the rupture itself also stops there.

  An appropriate physical condition describing the stable/unstable sliding across $L_{sc}$ might be whether the elastic stiffness $K$, as defined by $K=\delta f_{elastic}/\delta u$ which represents a change of the elastic force $f_{elastic}$ due to an infinitesimal slip $\delta u$ of the block, is greater/smaller than the frictional weakening rate, as defined by $\delta \phi/\delta u$ which represents a change of the friction force $\phi$ due to an infinitesimal slip of the block. Consider a hypothetical instantaneous process from the states ($u_i$, $v_i=0$, $\theta_i$) to ($u_i+\delta u_i$, $v_i=0$, $\theta_i+\delta \theta_i$). The aging law (\ref{aging}) entails the relation $\delta \theta_i=-\theta \delta u_i$. Then, the frictional-weakening rate is obtained as  $\frac{{\rm d}\phi}{{\rm d}u}=-b$. Meanwhile, the stiffness of the $L$-block system may be given by the smallest nonzero eigenvalue of the $L\times L$ matrix $K$ defined via the relation $(\delta f_{elastic,1}, \cdots, \delta f_{elastic,L})=K (\delta u_{1}, \cdots, \delta u_{L})$ as $K_{min}=2l^2\left( 1-\cos \frac{\pi}{L+1}\right) + 1$. Matching $K_{min}$ and $|\frac{{\rm d}\phi}{{\rm d}u}|$, the condition of the frictional instability is obtained as
\begin{equation}
L>L_{sc} = \frac{\pi}{\arccos \left( 1-\frac{b-1}{2l^2}\right)} - 1 .
\label{Lsc}
\end{equation}
The quasi-static initial phase is realizable in the BK model only when $L_{sc}$ is greater than the lattice spacing, {\it i.e.\/}, $L_{sc}>1$, or equivalently $b<b_c=2l^2+1$, yielding the condition of the weak frictional instability. We note that the formula (4) can also be derived from the linear stability analysis around the steady-state solution of the equation of motion, $v=v_{ss}=const.$ and $\theta=\theta_{ss}=1/v_{ss}$ along the line of Ref.\cite{Rice01}. 


 Since the continuum limit entails $l\rightarrow \infty$, the condition of the weak frictional instability $b<b_c=2l^2+1$ is always satisfied there. Hence, the continuum limit of the model always lies in the weak frictional instability regime accompanying the quasi-static nucleation process, corroborating Rice \cite{Rice}.

 In fig.2, we show for a typical large event in the weak frictional instability regime near the continuum limit  the time evolutions of the epicenter-block sliding velocity $v$, (a) in the initial phase, and (b) in the acceleration phase. In the initial phase, the sliding velocity $v$ stays very low up to the nucleation length $L_{sc}$, of order the pulling speed of the plate. In the acceleration phase, the block movement exhibits a prominent acceleration, being no longer quasi-static nor reversible, reaching the maximum around $L_c$ (this maximum point is used as our definition of $L_c$), then decreases sharply and finally stops.

\begin{figure}
\onefigure[scale=0.7]{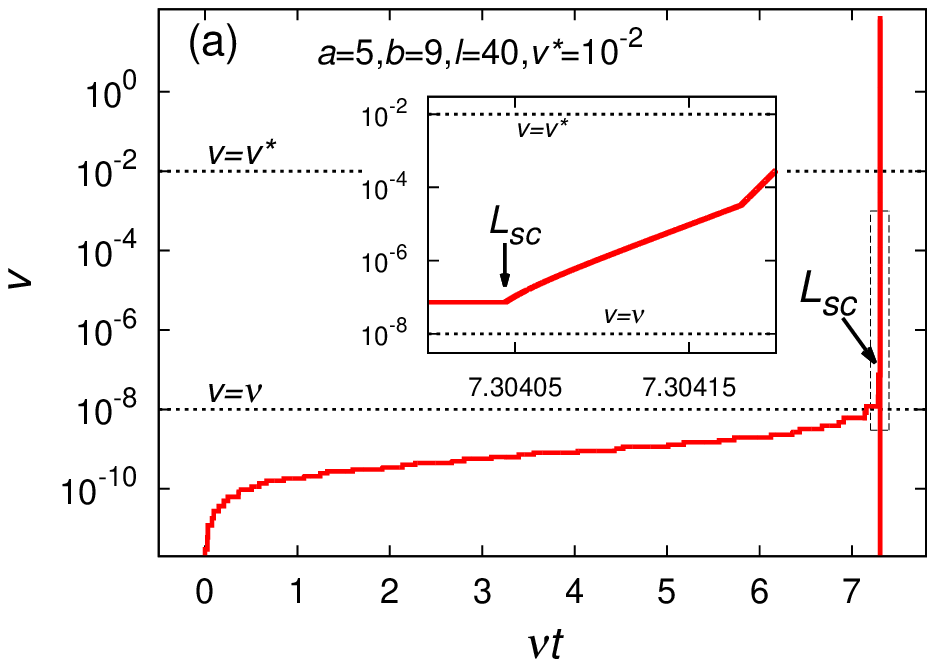}
\onefigure[scale=0.7]{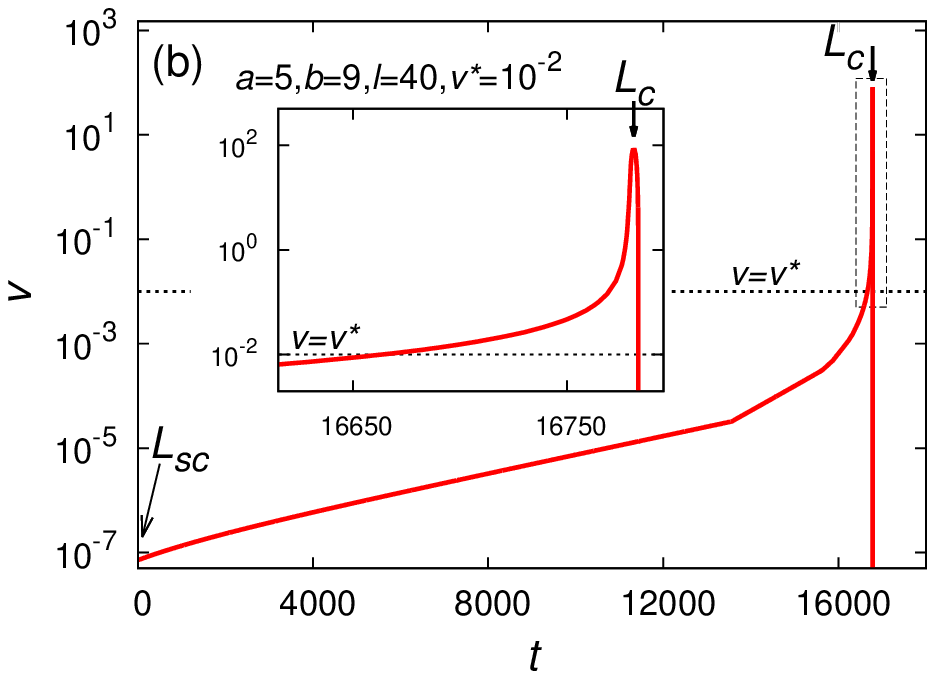}
\caption{
The time evolutions of the epicenter-block sliding velocity $v$, (a) in the initial phase, and (b) in the acceleration phase. The model parameters  are $a=5$, $b=9$, $c=1000$, $l=40$, $v^*=10^{-2}$ and $\nu=10^{-8}$. The time origin $t=0$ is taken at (a) the point where the epicenter begins to move, and (b) the point of $L=L_{sc}$. The arrows indicate the points of $L=L_{sc}$ and of $L=L_c$. The dotted horizontal lines represent the lines $v=\nu$ and $v=v^*$. The insets are magnified views. Note the abscissa is $\nu t$ in (a), but $t$ in (b).
}
\end{figure}

 Next, we investigate the statistical properties of the nucleation lengths $L_{sc}$ and $L_c$ as well as the duration times of each phase, averaged over many events (typically $10^4\sim 10^5$ events). Since the nucleation length $L_{sc}$ is determined only by the material parameters as in eq.(\ref{Lsc}), it cannot be used as an indicator of the size of the ensuing mainshock. What about $L_c$ ? We plot in fig.3(a) the computed mean $L_c$ normalized by the corresponding $L_{sc}$, $L_c/L_{sc}$, versus the final rupture-zone size $L_r$ for various choices of the model parameters in the weak frictional instability regime. The $b$-value is fixed to $b=9$ while the parameters $l$, $a$ and $v^*$ are varied. The data approximately collapse onto a common curve. Since $L_{sc}$ hardly depends on $a$ and $v^*$, this indicates that $L_c$ is insensitive to $a$ and $v^*$, while its $l$-dependence is the same as that of $L_{sc}$. One also sees that large events tend to be independent of $L_r$, implying that one cannot predict the size of the upcoming mainshock even with the information of $L_c$. We examine the $b$-dependence of $L_c/L_{sc}\equiv r$, to find that it increases linearly with $b$ as $r(b)\simeq 0.1b+4.4$ in the weak frictional instability regime $b<b_c$.

\begin{figure}
\onefigure[scale=0.7]{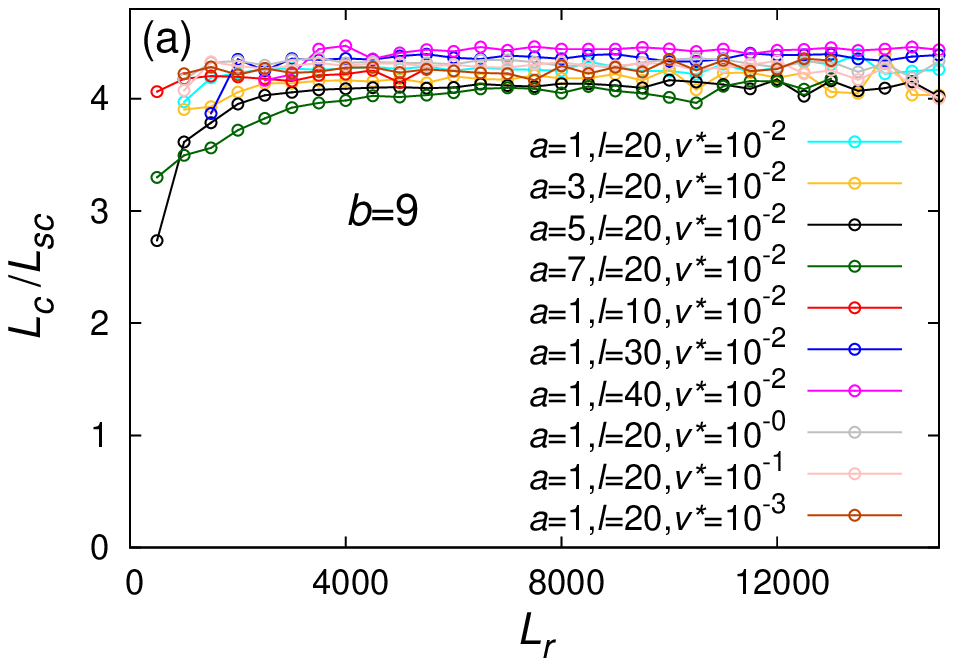}
\onefigure[scale=0.7]{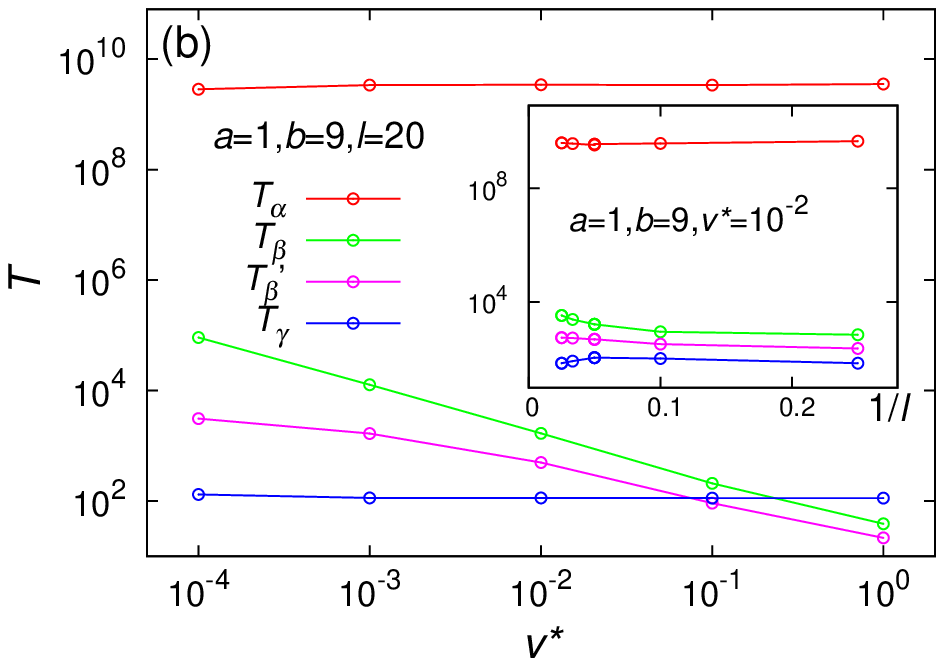}
\caption{
(Color online) (a) The mean nucleation length $L_c$ divided by $L_{sc}$, $L_c/L_{sc}$, plotted versus the rupture-zone size $L_r$ for various parameter sets {$a$, $l$, $v^*$} in the weak frictional instability regime. The other parameters are fixed to $b=9$, $c=1000$ and $\nu=10^{-8}$. (b) The mean duration times plotted versus the crossover velocity $v^*$ with $l=20$ (main panel), and versus the inverse stiffness parameter $1/l$ with $v^*=10^{-2}$ (inset). The other parameters are $a=1$, $b=9$, $c=1000$ and  $\nu=10^{-8}$. Both in (a) and (b), the average is made over $10^4\sim 10^5$ events for each data point.
}
\end{figure}

 The continuum limit of $L_{sc}$ in the dimensionless form $\tilde L_{sc}\equiv \lim_{d\rightarrow 0} L_{sc}d=\lim_{l\rightarrow \infty} \frac{L_{sc}}{l}$ is obtained as $\tilde L_{sc} =\pi/\sqrt{b-1}$. Reviving the normalization units above, we get the dimensionful nucleation length in the continuum limit, $L_{sc}^\times$, as
\begin{eqnarray}
L_{sc}^\times = \pi \sqrt{ \frac{G{\mathcal L}\omega}{\sigma_n v_s B} } \frac{v_s}{\omega} ,
\label{Lsc2}
\end{eqnarray}
 for $b>>1$. If we substitute the parameter values $\mathcal{L} \simeq 1$ [cm], $B\simeq 10^{-2}$ and $\frac{v_s}{\omega} \simeq 2$ [km], we get $L_{sc}^\times$ several kilometers. Concerning $L_c$, since the ratio $r=L_c/L_{sc}$ turns out to be hardly dependent on $l$ in the weak frictional instability regime, the dimensionful nucleation length in the continuum limit $L_c^\times$ is given by $L_c^\times = r(b) L_{sc}^\times\simeq (0.1b+4.4)L_{sc}^\times$, where $b=\frac{\sigma_n v_s}{G \omega{\mathcal L}}B$ is a number characterizing the fault interface.

 Next, we consider the duration times of each stage of the nucleation process, including that of the initial phase $T_\alpha$ ($L< L_{sc}$), of the acceleration phase $T_\beta$ ($L_{sc}<L<L_c$), and of the high-speed rupture phase $T_\gamma$ ($L>L_c$). The ultimate utility of nucleation phenomena may be forecasting the upcoming mainshock. Practical detection, if any, would become possible in the acceleration phase. Since the system has already been beyond the ``no-return'' point, a mainshock should already be ``deterministic''. The remaining problem is how much time is left there. We tentatively set the detectable sliding velocity of the nucleus motion $v =10^{-4}=10^4\nu$ which corresponds in real unit to $\simeq 10^{-2}$ [mm/sec]. Then, the time interval between the point of $v=10^{-4}$ and the point of $L=L_c$ (the onset of a mainshock) is denoted by $T_\beta^\prime$. This $T_\beta^\prime$ would give a realistic measure of the remaining time available for a mainshock forecast.

 Among various duration times, though the duration time of a mainshock $T_\gamma$ naturally increases with the size of a mainshock $L_r$, $T_\alpha$, $T_\beta$, and $T_\beta^\prime$ hardly depend on $L_r$ except for smaller events. This means that it is again hard to predict the size of the mainshock based on the duration time of the nucleation process. We also examine the dependence of these duration times on the model parameters, to find that they are hardly dependent on $a$, $b$, $l$, but $T_\beta$ and $T_\beta^\prime$ sensitively depend on the friction crossover velocity $v^*$. In fig.3(b), we plot the mean duration times versus $v^*$ (main panel), and versus $1/l$ (inset). For $v^*=10^{-4}$, $T_\beta$ is greater than $T_\gamma$ by factor of 700, while $T_\beta ^\prime$ by factor of 20. The $1/l$-dependence of these duration times shown in the inset turns out to be rather weak. Then, reviving the normalization units and substituting the typical parameter values, we estimate, for $v^*=10^{-4}$, $T_\alpha\simeq 10^2$ [year], $T_\beta\simeq 1$ [day], $T_\beta^\prime\simeq 1$ [hour] and $T_\gamma\simeq 1\sim 2$ [min] in the continuum limit. For smaller $v^*$, $T_\beta^{\prime}$ could be even longer, but taking the data for $v^*\leq 10^{-5}$ is beyond our present computational capability. We conclude that the remaining time available for a mainshock forecast could be longer than the mainshock duration time by one or two orders of magnitude, but perhaps not much longer than that. Of course, since the reliability of the 1D BK model in connection with real seismicity may be limited at the quantitative level, these estimates should be taken only as rough measures.

 In summary, we studied the properties of the earthquake nucleation process as a precursor of a mainshock both  numerically and analytically on the basis of the BK model obeying the RSF law, and found that this simplified model successfully reproduced various features of the expected earthquake nucleation process. 
 
 The authors are thankful to T. Okubo, T. Hatano, N. Kato, T. Uchide and S. Ito for useful discussion. They are also thankful to one of the referees for pointing out the possible connection of the BK model to the LVFZ, and bringing Ref.\cite{Huang} to the authors' attention. This study was supported by Grant-in-Aid for Scientific Research on Priority Areas 21540385. We thank ISSP, Tokyo University for providing us with the CPU time.

\end{document}